# Frequency and Impact of Technical Debt Characteristics in Companies Producing Mechatronic Products

Fandi Bi, Birgit Vogel-Heuser, Litong Xu
Department of Mechanical Engineering
Technische Universität München
Garching b. München, Germany
{fandi.bi, vogel-heuser, litong.xu}@tum.de

## ABSTRACT

Complexity of products, volatility in global markets, and the increasingly rapid pace of innovations may make it difficult to know how to approach challenging situations in mechatronic design and production. Technical Debt (TD) is a metaphor that describes the practical bargain of exchanging short-term benefits for long-term negative consequences. Oftentimes, the scope and impact of TD, as well as the cost of corrective measures, are underestimated. Especially for mechatronic teams in the mechanical, electrical, and software disciplines, the adverse interdisciplinary ripple effects of TD incidents are passed on throughout the life cycle. The analysis of the first comprehensive survey showed that not only do the TD types differ in cross-disciplinary comparisons, but different characteristics can also be observed depending on whether a discipline is studied in isolation or in combination with others. To validate the study results and to report on a general consciousness of TD in the disciplines, this follow-up study involves 15 of the 50 experts of the predecessor study and reflects the frequency and impact of technical debt in industrial experts' daily work using a questionnaire. These experts rate 14 TD types, 47 TD causes, and 33 TD symptoms in terms of their frequency and impact. Detailed analyses reveal consistent results for the most frequent TD types and causes, yet they show divergent characteristics in a profound exploration of discipline-specific phenomena. Thus, this study has the potential to set the foundations for future automated TD identification analyses in mechatronics.

## 1 Introduction

The Technical Debt (TD) metaphor was coined by Cunningham [1], and has evolved with the Dagstuhl 16162 definition [2]. According to Dagstuhl 16162, TD is defined as "a design or implementation construct that is expedient in the short term but sets up a technical context that can make a future change more costly or impossible "[2]. Often, the scope and long-term damage of TD to the system are underestimated. This is especially the case when multiple disciplines are involved, which do not see how their decisions affect the others, e.g., management, different mechatronic disciplines, and or purchasing department. The dependencies between different disciplines in combination with the shortened and asynchronous innovation cycles of mechatronic products create unpredictable TD that is difficult to manage. The prior study "*TDinMechatronics*", involving 50 expert interviews from 21 companies, analyzed the current state of TD in the industry [3]. The 94 collected TD incidents reflect TD in the life cycle of mechatronic products and focus on the domain-specific view of TD types. Although *TDinMechatronics* exposed new TD characteristics, the results gathered have not yet been validated, to directly correlate to the number of TD incidents in the particular experts' workplace. The experts presented TD incidents that, in their view, were most relevant and significant. Therefore, for this study, we gather the TD types, TD causes, and TD symptoms identified in the predecessor study [3] and enlarge them by adding the most frequent ones of software engineering [4, 5]. We involve the same group of experts and focus on their ratings, reflecting the overall situation of TD in their discipline and company.

We investigate the following research questions (RQs):

RQ 1   How do the results on *TDinMechatronics* represent the actual state of TD to experts?

RQ 2   What are the TD types, TD causes, and TD symptoms with the most considerable frequency and most significant impact?

RQ 3   How does the discipline of the expert affect the frequency and impact of TD characteristics?

RQ 4   Which corporate and organizational structures encourage the avoidance of initiating TD?

In the next section, we provide the background and related work of TD in software engineering and mechatronics, while elaborating on the potential benefit to practitioners in the field of mechatronics. Section 3 presents the applied research methodology. In Section 4 and 5, we conclude the findings, discuss, and compare the results to previous related studies.

## 2 Background and Related Work

This section introduces empirical studies from software engineering that investigate TD frequency and impact, and presents the status of TD research in mechatronics.

## 2.1 TD in Software Engineering

TD research emerged in 1992 and has increasingly gained attention in recent years in software engineering [1, 6]. According to the classification of Avgeriou et al. [2], TD is composed of different TD items. TD items can be interdependent, they can be triggered by TD causes, exhibit TD symptoms, and lead to TD consequences. Li et al. [6] present ten TD types for software systems, with Code TD, Architectural TD, Test TD, and Design TD being the most commonly studied ones. Ampatzoglou et al. [7] performed a case study involving seven embedded systems industries. The findings indicate that some quality attributes such as functionality, reliability, and performance are being given higher priority compared to TD management (TDM). At the same time, the most recurring TD types in the embedded software industry are Test TD, Architecture TD, and Code TD. Finally, developers acknowledge the need for low TD on components that are expected to have a longer lifetime, compared to more short-lived ones. To support development teams in the task of TDM in software projects, Rios et al. [4, 5] presented diagrams composed of cross-company data, which allow development teams to have a broad overview of eventual TD causes and TD effects. The frequency of up to 105 TD causes and 85 TD symptoms were summarized in Ishikawa diagrams. By comparing their experience on previous projects and the content of the diagrams, developers can calculate possible upcoming TD causes and TD effects. Participants of the study agreed that, by using the proposed diagrams, they gain agility, productivity, performance, and effectiveness for the identification of TD.

## 2.1 TD in Mechatronics

Not only does the TD metaphor apply to software-intensive systems, but TD was also identified in mechatronic systems with new characteristics [8, 9]. TD occurs in all three mechatronic disciplines, namely mechanics, electrics/electronics, and software, as well as in some of the additional phases of the life cycle, e.g., manufacturing, assembly, commissioning, and handover to the customer, and service [8]. Similarly, the follow-up study exemplifies the TD types of Requirements, Architectural, Design, Test, Infrastructure, and Documentation TD according to Li et al. [6, 9]. As causes for TD in production systems, inconsistencies, information asymmetry between mechatronic disciplines, misinterpretation of technical and customer requirements due to insufficient specifications, and lack of budget for maintenance are identified as examples [8]. The findings of Besker et al. [10] reveal that TD is a big challenge in mechatronic companies. Furthermore, these companies want to understand if TD is merely related to software, or if it is rooted in other disciplines as well. Their results suggest that there is a difference between which TD types generate more effort, and this is related to the combination of software and hardware development. *TDinMechatronics*, with 50 industrial experts from ten industries [3], shows the divergent distribution of TD types between mechatronic systems and software systems. In the first three phases of the life cycle, the most frequently identified TD types are Requirements TD, Architectural TD, and Design TD, with Infrastructure TD mentioned as the most common TD type excluded from the life cycle. The preliminary work on TD in mechatronic production systems was able to exemplarily identify all TD types from software systems, except Build TD [3, 6]. Due to the further phases of the life cycle and the extra components of mechatronic systems, additional TD characteristics arise for mechatronic systems. Quantifiable and validated statements that represent the frequency and impact of TD on mechatronics and the consideration of causality between TD elements (types, causes, symptoms) are missing [3].

## 2.3 Research Gap

TD characteristics that receive maximum attention have already been identified in software and mechatronic systems. Elaboration on their frequency and impact is now to be expected. This study focuses on investigating the validity of the previous expert interview results [3] for a deeper understanding of TD in mechatronics. Furthermore, the avoidance of TD is to be examined, based on whether it depends on distinct characteristics of the product or the company, e.g., output quantity, external conditions/requirements, or internal regulatory mechanisms. The potential benefit of this work is the report of the real impact of TD characteristics on practitioners in the field of mechatronics, by detecting and collecting qualitative and quantitative statements about TD types, causes and symptoms in companies producing mechatronic products.

## 3 Research Methodology

In this section, we present the study design, the data acquisition, the procedure and protocols, and the method applied in the data analysis. To elaborate on the semistructured expert interviews of *TDinMechatronics*, we selected questionnaires to ensure consistent data collection in this study [11].

### 3.1 Study Design

The results of the *TDinMechatronics* and the questionnaire were sent to the experts who originally participated in the study. A number of 15 participants got back to us within 14 days. Due to the large number of TD causes and symptoms on evaluation, i.e. questions of over 74 rating gaps (37 TD causes for rating frequency and impact), our study design is strongly limited. We decided to offer two hand-in options, a digital and an analogue one, and selected the excel forms due to their simplicity. The format of Excel forms offers a better overview of the questions and the opportunity to print out the questionnaire and return the manually filled in form by post. To foster comparability, for all closed questions regarding the frequency and impact evaluation, we selected the numerical coding using the Likert scale [12] (numbers between "1-6"). In case an expert cannot relate to a particular TD characteristic, a rating of "0 "is possible.

### 3.2 Questionnaire Structure

The interview questionnaire features a mix of six open and eight closed questions. Three open questions in the first section relate to the feedback and lessons learned from the previous study. The second section, with three subsections, evaluates the frequency and impact of TD Types, TD causes, and TD symptoms. Here, we gathered all TD types of *TDinMechatronics*, resulting in nine types from TD in software engineering [6] (9/10, except Build TD) and four types from TD in mechatronics (Start Up TD, Maintenance TD, and Variants TD, Manufacturing TD) [3, 9]. TD aspects concerning the subject of work production planning are introduced as Production Planning TD [13], from on-site changes by technicians to late customer requests, often result in variants existing in parallel that may cause refactoring and may require a change of components in different types of plants.

Furthermore, regarding the TD causes and TD symptoms, we adopted those identified options with more than three occurrences, and enlarged them with the causes and symptoms of Rios et al. [4, 5], having more than a 5 % occurrence rate. Following the clusters chosen in [4], we categorized the options into eight TD cause clusters and six TD symptom clusters. In the third section of the questionnaire, the TD causes and TD symptoms with a gray cell color[2] indicate their origin from Rios et al. [4]. The third section of the questionnaire with three open questions investigates the consequences of TD and their corresponding tracking mechanisms. The final section, with five closed questions, assesses the avoidance of TD. Experts evaluate five hypotheses for avoiding TD.

### 3.3  Data-Preprocessing and Analysis

The data-preprocessing phase starts once the dataset is fixed. Considering the closed questions are made up of numbers in the range [0;1, 6], they do not require any normalizing. In regard to the open questions, we applied the pattern coding method [14]. Furthermore, the encoding was discussed among the authors to reduce misunderstandings. The paper-and-pencil hand-ins were manually converted into digital excel forms. For the digital excel forms, we created a semi-automated data transfer of the multiple received questionnaires into one excel table for data analysis. This offers the correlation analysis between TD characteristics, as well as the comparison of previous results to verify the outcomes of *TDinMechatronics*. We used Tableau Desktop[3] to analyze and visualize the results.

## 4  Questionnaire Results

This section summarizes the results of the 15 questionnaires we have received. Firstly, we introduce statistics describing the demographics of the participants, proceeding on to the related quantitative data to address each research question.

### 4.1  Descriptive Statistics and Response to TDinMechatronics Study

*4.1.1 Descriptive Statistics.* Overall, we collected 15 expert responses from the 50 experts who participated in the initial study (cp. Tab. 1). We adopted the same classification as in the previous study. A leading position refers to the position of group manager or higher. This study quantifies their work experience using the time spent working in mechatronics with a minimum of two years required. In the sample, 80 % of the experts had working experience in mechatronics of at least ten years. As all experts involved participated in the previous expert interviews, their understanding of TD was coined according to the Dagstuhl 16162 definition [2].

Tab 1. Key information of surveyed experts and companies of this study; *Interviewee ref. [3]; **Discipline: ME = Mechanical Engineering, EE = Electrical/Electronical Engineering; SE = Software Engineering; ***Company size: L = Large (over 250 employees); M = Medium (50-249 employees); n = 15

| Interviewee* | Working experience in years | Discipline** ME (27%), EE (20 %), SE (33%), EE+SE (7%), ME+EE+SE (13%) | Leading position (67 %) | Company size*** |
|---|---|---|---|---|
| 5 | 2-9 | EE | no | L |
| 6 | 10-19 | EE | no | L |
| 7 | 20+ | ME | no | L |
| 13 | 10-19 | SE | yes | L |
| 14 | 20+ | SE | yes | L |
| 24 | 2-9 | EE, SE | yes | L |
| 26 | 20+ | ME, EE, SE | yes | L |
| 27 | 2-9 | SE | yes | L |
| 28 | 10-19 | SE | yes | L |
| 29 | 2-9 | EE | yes | L |
| 30 | 10-19 | SE | no | L |
| 31 | 2-9 | ME, EE, SE | yes | L |
| 35 | 10-19 | ME | yes | M |
| 45 | 2-9 | ME | no | L |
| 48 | 10-19 | ME | yes | L |

*4.1.2 Response to TDinMechatronics Study.* A significant number of our experts (42 %) state that the quantification of the TD types was crucial, and that it was critical to find out whether Requirements TD, Architecture TD, Infrastructure TD, or Design TD occur most frequently in mechatronics development. A majority of the experts (64 %) did not expect that most of the TD incidents occur at the beginning of the life cycle, Requirements TD and Infrastructure TD in particular. Furthermore, it was surprising that in the implementation of a project, less debt arises (Test TD, Code TD and Versioning TD). Given the fact that the frequency of occurrence of the TD types peaks at the beginning of the life cycle of mechatronic development and decreases over the process, a further number of the experts (33%) believes that an exact planning phase and precise execution of the planning are irreplaceable. According to the leverage effect in product life cycles, small savings, compromises, and accelerations in the early stages may generate significant consequences in the later stages.

---

[2] Questionnaire (https://mediatum.ub.tum.de/1592082)

[3] https://www.tableau.com/products/desktop (Version 2020.4.2)

Especially in preceding processes, decisions are to be made strictly according to systematic routines. Furthermore, two experts (17%) state that their renewed perception on or confirmation of the importance of interdisciplinary engineering, given the spread of TD across different disciplines, is the most critical information they gained from the *TDinMechatronics* study. The complex and interdisciplinary nature of TD in mechatronics and the high share of TD involving the management are surprising. Thus, one expert states that requirements engineering should be designed in such a way that all disciplines are well represented, know the goal, and are aware of the way towards the goal. The interdisciplinary communication plays a central role in terms of this issue [15]. One expert states that it is surprising and regrettable that TD can be caused by missing communication. A clear majority (60%) of the experts expect further investigations focusing on criteria for identifying underlying TD incidents and the evaluation of its consequential costs using, e.g., impact analyses.

## 4.2 Frequency and Impact of TD Characteristics

In this questionnaire, the experts rated a total of 14 TD types, 47 TD causes, and 33 TD symptoms in terms of their frequency and impact from "0 "(none) to "1 "(lowest) to "6 "(highest). Besides the analyses of the TD characteristics, we compared these 15 experts' answers to the results of their initial expert interview.

*4.2.1 TD Types – Frequency and Impact.* The statistics reveal that Requirements TD (average frequency = 4.1, average impact = 4.9) is the most dominant TD type regarding the frequency and impact in mechatronics (cp. Tab. 2). Followed by Infrastructure TD and Variants TD, Architecture TD and Design TD prove to be the two remaining TD types with above-average frequency. On the other hand, Defect TD, Versioning TD, and Code TD are the most uncommon TD types according to the judgment of the experts. In terms of frequency, Variants TD, Requirements TD, Code TD, and Documentation TD show the widest quartiles. Thus, the opinions of the polled experts on these TD types are the most unevenly distributed. The expert opinions on the impact of Variants TD, Infrastructure TD, Start Up TD, and Test TD change sharply. In Design TD, the opinions of experts are relatively consistent in terms of frequency and impact. The column " Δ average-impact "shows that in 3/14 TD types, the impact receives higher scores than the frequency. Documentation TD occurs relatively often, but its impact is comparatively small, while Manufacturing TD, Architecture TD, and Test TD cause significant damage once they arise.

Fig. 1 shows the disciplinary comparison between the initial study and the current one for TD types. As the groups of interdisciplinary experts have only one (EE+SE) and two (ME+EE+SE) representatives, we focus on the analysis of the other three clusters. For mechanical engineering (ME), Infrastructure TD, Requirements TD, and Documentation TD occur most frequently based on the previous results. However, in this study, these TD types received rather little attention from the mechanical engineering experts, who rated those TD types as average in frequency. Furthermore, the mechanical engineering experts classified Design TD as both most frequent and impactful. However, in the previous study, the same group of experts did not mention Design TD at all. Instead, merely a medium-sized group of mechanical engineers (cp. Fig. 1, 11 mentions, ranked 4[th]), who did not participate in this study, mentioned Design TD. The same applies to Maintenance TD. In the discipline of electrical/electronical engineering (EE), Infrastructure TD, Design TD, and Documentation TD were identified as frequently reappearing TD types by all experts, both in this study and in the previous study. However, these TD types received merely average ratings in terms of their frequency in this study. Meanwhile, Start Up TD was not mentioned in *TDinMechatronics* at all. Instead, in this study, Start Up TD receives a high frequency rating. In the discipline of software engineering (SE), Requirements TD remains the single most crucial TD type in both studies, receiving both the highest total occurences. Code TD, Start Up TD, and Versioning TD were rated as relatively uncommon and harmless. In the *TDinMechatronics* study, Variants TD was merely mentioned once in 220 TD items, whereas in this questionnaire, this new TD type Variants TD gained the second-highest score in terms of both frequency and impact.

Tab 2. TD type frequency and impact; [a)] *TDinMechatronics*; [b)] results of this study

| TD type | TD type in TD-inMechatronics [% of 220 TD items][a)] | Avg. Frequency (avg. 3.1)[b)] | Avg. Impact (avg. 3.5)[b)] | Δ average -impact [b)] |
|---|---|---|---|---|
| Requiremmments TD | 19 | 4.1 | 4.9 | -0.8 |
| Variants TD | 4 | 3.7 | 4.4 | -0.7 |
| Infrastructure TD | 28 | 3.7 | 4.3 | -0.6 |
| Architecure TD | 8 | 3.5 | 4.4 | -0.9 |
| Design TD | 12 | 3.4 | 4.1 | -0.7 |
| Test TD | 6 | 3.0 | 3.8 | -0.8 |
| Start Up TD | 1 | 3.0 | 3.3 | -0.3 |
| Documentation TD | 8 | 3.0 | 2.4 | +0.6 |
| Maintenance TD | 4 | 2.9 | 3.2 | -0.3 |
| Manufacturing TD | 0 | 2.8 | 3.9 | -1.1 |
| Work Preparation TD | 0 | 2.8 | 3.1 | -0.3 |
| Code TD | 2 | 2.8 | 2.6 | +0.2 |
| Versioning TD | 4 | 2.6 | 2.4 | +0.2 |
| Defect TD | 5 | 2.4 | 2.8 | -0.4 |

4.2.2 *TD Causes – Frequency and Impact.* For mechanical engineering, "Save Costs "is a significant TD cause receiving frequent mentions in both studies. In *TDinMechatronics*, "Decision Without Specific Knowledge Though Needed ", "Different KPI In Departments ", and "Lack Of Communication "were repeatedly described by mechanical engineering, but rated as below-average in this study (cp. Fig. 2).

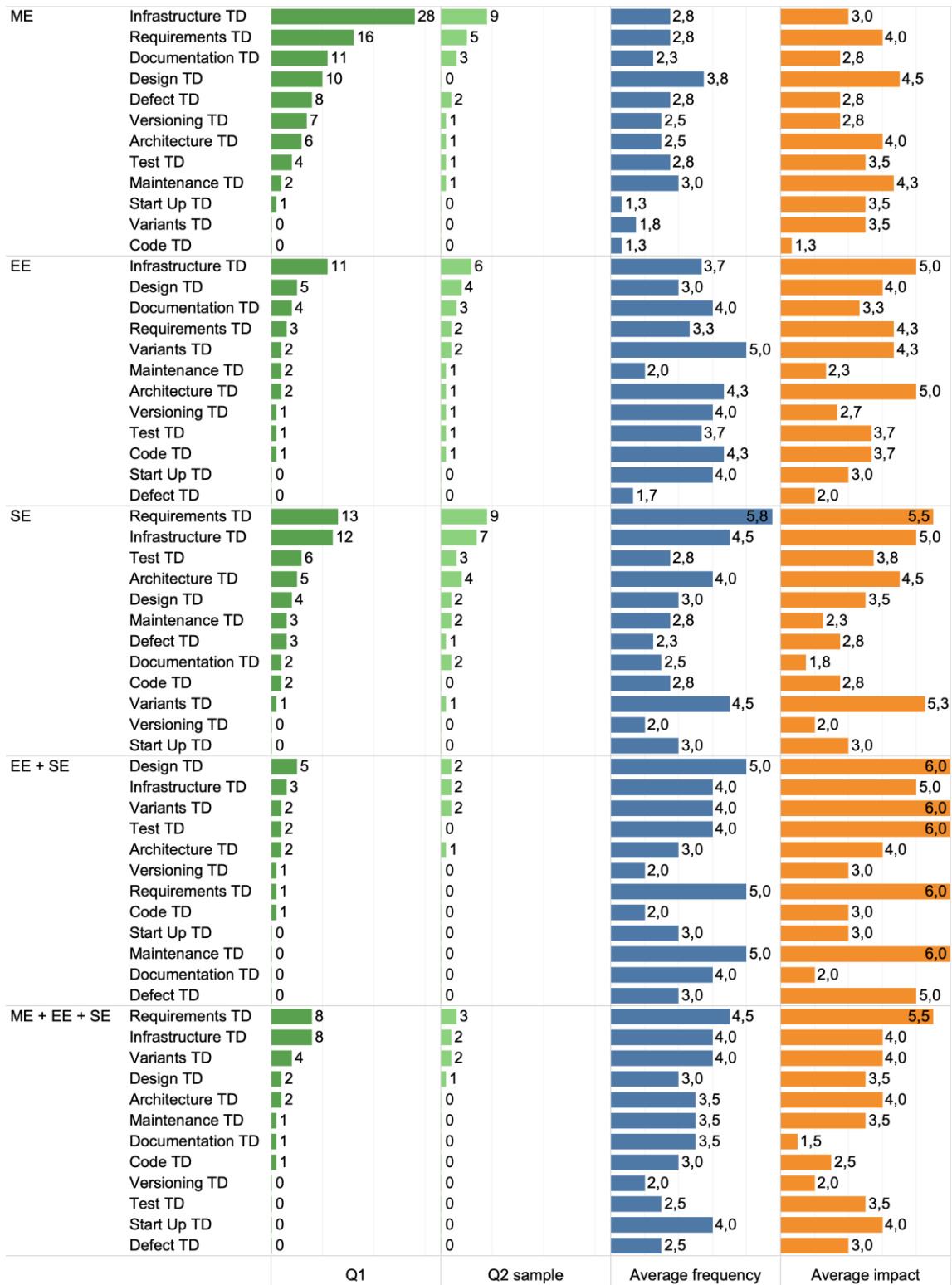

**Fig. 1** Comparison between the number of TD types in TDinMechatronics (Q1), number of TD types in *TDinMechatronics* mentioned by the same expert sample as in this study (Q2 sample), average frequency, and average impact in this study, categorized by mechatronic disciplines; ME = Mechanical Engineering, EE = Electrical/Electronical Engineering; SE = Software Engineering

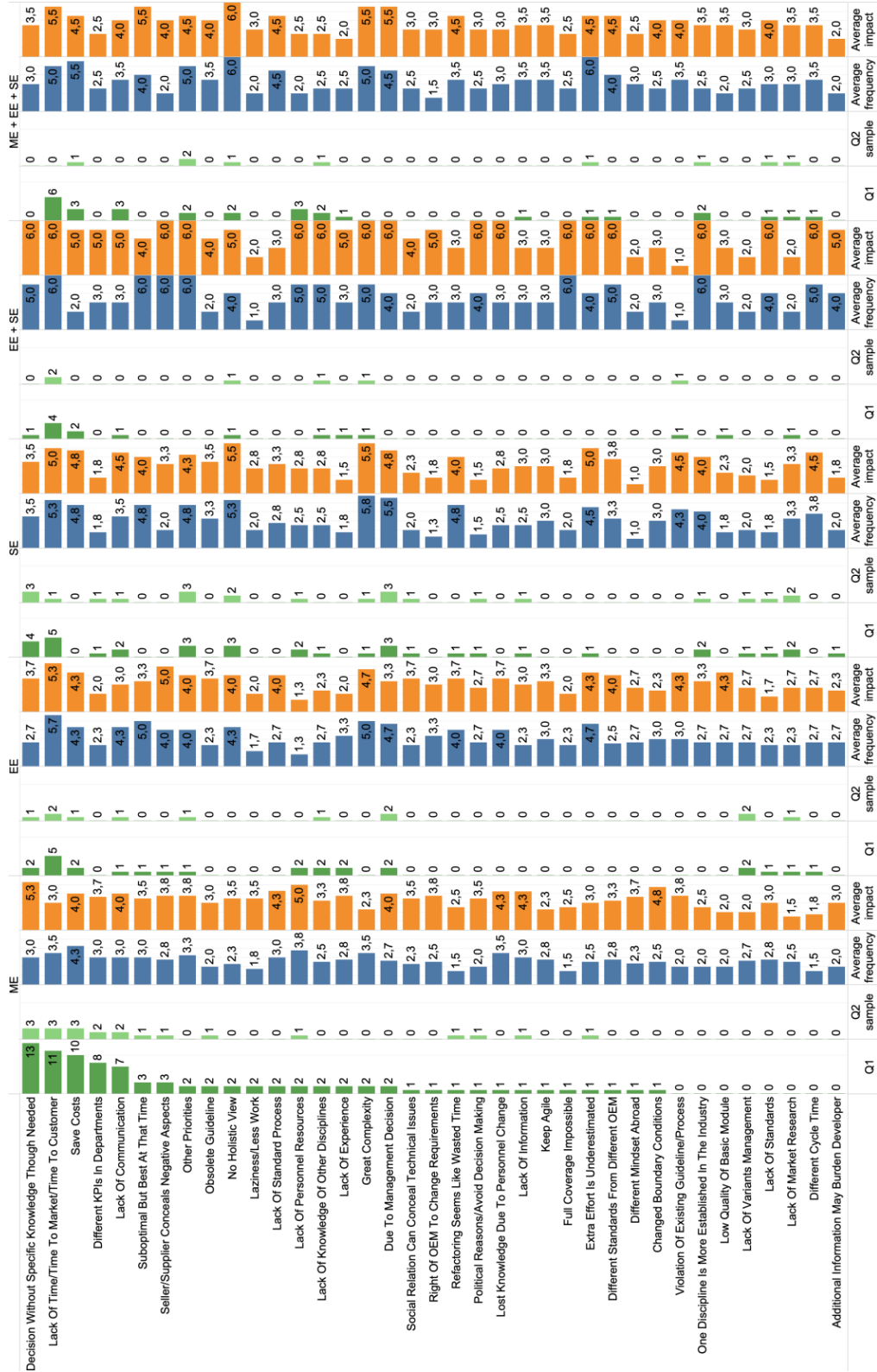

Fig. 2 Comparison between the number of TD causes in (Q1), the number of TD causes in *TDinMechatronics* mentioned by the same expert sample as in this study (Q2 expert sample), average frequency, and average impact in this study, categorized by mechatronic disciplines; ME = Mechanical Engineering, EE = Electrical/Electronical Engineering; SE = Software Engineering

Electrical engineers often mentioned "Lack of Time/Time To Market/Time To Customer "as a TD cause in previous interviews. In this study, said TD cause was rated accordingly as the most frequent TD cause. However, the evaluations of the remaining TD causes show fewer similarities with previous results. "Great Complexity", "Extra Effort Is Underestimated", and "Refactoring Seems Like Wasted Time" instead are marked as highly frequent TD, although they were not mentioned in the previous study by electrical engineers. As for software engineers, "Great Complexity "is the most frequent and impactful TD cause here, yet it was barely mentioned before. The same applies to the TD causes "Suboptimal But Best At That Time "and "Save Costs". Meanwhile, TD causes, e.g., "Lack Of Time/Time To Market/Time To Customer", "Due To Management Decision", and "No Holistic View "often appeared in both studies in the category of software engineering. The significance of interdisciplinary engineering is once again demonstrated.

*4.2.3 TD Symptoms – Frequency and Impact.* Overall, "Bad Performance/Fast Wear Or Tear "is the most critical TD symptom for mechanical engineering in both studies (cp. Fig. 3). "Broken Part/Error/Defect "and "Feeling Of Sub-optimal Solution "were repeatedly mentioned in the previous study but show low rates here. However, "Rework Needed "is clustered as a frequent TD symptom to be detected in this questionnaire, but it was not mentioned before. "Customer Complaint", which is a highly impactful TD symptom according to these results, was not mentioned in the previous study either. For electrical engineering, "Feeling Of Sub-optimal Solution/Known Risk "was the most mentioned TD symptom in the previous study that also reports high frequency in this study. However, the two TD symptoms, "No Transparent Decision Finding Process" and "Rework Needed "that are also classified as very frequent in this study, were never mentioned in the previous study by the electrical engineering experts. The software engineers often mentioned "Bad Performance/Fast Wear Or Tear "in their TD incidents in the interviews, but the classification in this study shows only average frequency. Instead, experts stress that the TD symptom "Inconsistencies "is highly frequent with large impact, in spite of the fact that it did not appear in the previous study.

### 4.3 Structures Encouraging the Avoidance of TD Initiation by Investigating the TD Consequences

Corporate and organizational structures can reflect the way TD are handled in a company. By investigating five aspects regarding different production circumstances, we aim to identify patterns that encourage or discourage the avoidance of initiating TD. Experts rated five circumstances on the output quantity, market conditions, mechatronic development, complexity of variants, and the TD cause tracking from 1 (least prevention) to 6 (largest prevention), non of the experts selected 0 (cannot relate to TD characteristic) (cp. Fig. 4). With an average of 4.5, "Larger Output Quantity "has the most significant positive influence on the experts in terms of avoiding TD. Furthermore, 47 % of all experts voted for the highest prevention potential (6). "Uncertain Market Conditions "leads to the second-highest caution in the generation of TD with an average score of 4.1. Exactly 60 % of all experts share the opinion that above-average attention needs to be paid to avoiding TD when the markets are volatile (blue areas, 3-6). It is worth noticing that none of the experts agreed on the least TD prevention (1) in this case. This is unique among all production circumstances. Ranked third, "Mechatronic

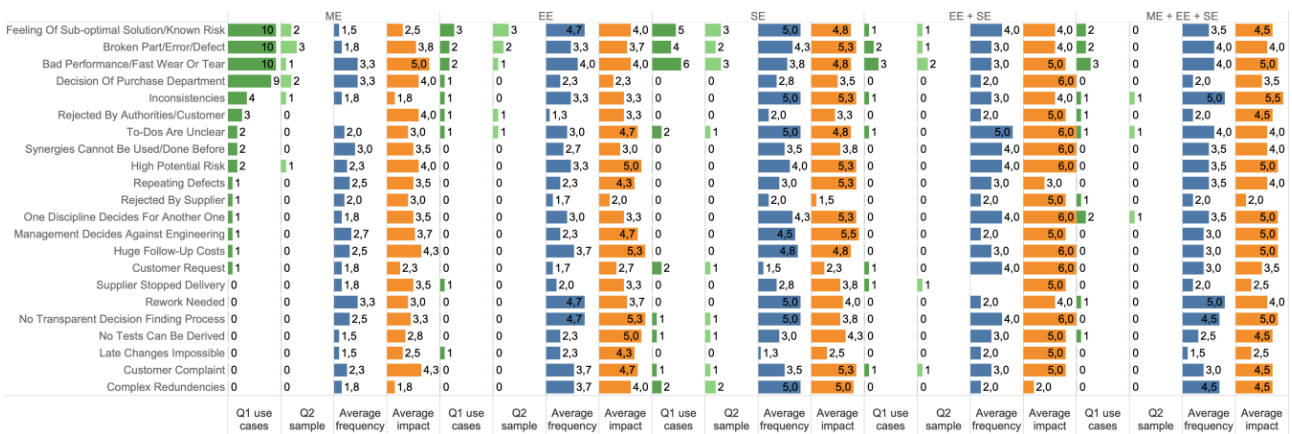

**Fig. 3** Comparison between the number of TD symptoms in TDinMechatronics (Q1), the number of TD causes in TDinMechatronics mentioned by the same expert sample as in this study (Q2 expert sample), average frequency, and average impact in this study categorized by mechatronic disciplines; ME = Mechanical Engineering, EE = Electrical/Electronical Engineering; SE = Software Engineering

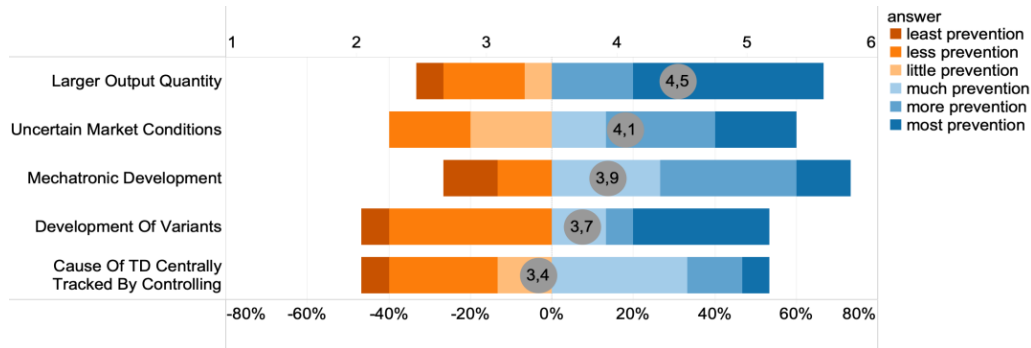

**Fig. 4 Structures encouraging the avoidance of TD incidents' initiation in mechatronics**

Development" received an average score of 3.9. Up to 73 % of all respondents state that in this interdisciplinary development environment, TD is to be prevented (3-6). This share of experts giving a positive correlation is the highest among all, although the share of "most prevention "is comparatively low at 13 %. With an average of 3.7, the number and complexity of variants summarized in "Development of Variants "has a comparatively small, yet positive effect on the avoidance of TD. The distribution of experts splits in half, into above and below average opinions. Whether the "Cause Of TD is Centrally Tracked By Controlling "or not, the level of avoidance of TD barely changes. With an average of 3.4, a centrally tracked mechanism leads to a slight decrease in the efforts of TD prevention management compared to a locally tracked one.

## 5    Discussion, Comparison and Study's Validity

This section interprets the study findings and compares our results to previous studies. Furthermore, Yin's [16] four aspects of validity addressing the threats, which may jeopardize the validity of the results, are addressed.

### 5.1    Discussion of the Study Findings

Regarding the comparison of the conclusions reached in *TDinMechatronics* and this questionnaire, the conclusions show predominant similarities. Those TD characteristics (TD type, TD cause, TD symptom), which were often mentioned before, also received high ratings in terms of their frequency and impact in this study. However, some TD characteristics (Variants TD, Maintenance TD, Start Up TD) that were merely occasionally mentioned in *TDinMechatronics* were rated as highly frequent with a large impact by the experts in the questionnaire. Hence, *TDinMechatronics* does not give an entire picture of all TD incidents and TD characteristics occurring in companies with mechatronic products. Due to the limited interview time of between 30 and 45 min, experts were able to present 1-8 TD incidents. Hence, they picked the ones that were the most relevant to them. This leads to the fact, that on the one hand, there are some peaks in selected TD types (e.g. Requirements TD, Infrastructure TD). Most probably, these TD types actively expose TD incidents in the system. In the same life cycle, or even shortly after the initiation, the expert becomes aware of them. We can consider them as "active TD types ". On the other hand, those TD characteristics (e.g., Variants TD) that may not directly lead to a TD incident outbreak exist passively in the engineering and may "survive "more than one cycle and are less prominent. Therefore, experts may not directly link these appearances to TD.  Yet, in case they are remembered when seeing the option on the questionnaire, experts rerecord their high frequency and impact. These TD characteristics may be considered as "passive TD types". The research of passive TD types, of how long they can remain uncovered in a system, might be of great interest.

Mechatronic disciplines have a significant impact on the perception of TD characteristics. Due to their unique position in interdisciplinary development, mechanical engineers gave TD characteristics the lowest ratings in terms of their frequency and impact. They are notably less confronted with TD incidents regarding adjustment, optimization, interdisciplinary development, and lack of infrastructure. On the contrary, software engineers working at the final stage of interdisciplinary development often face failures in decision-making, redundancies, and missing cooperation. The opinions of electrical engineers are located in the area between mechanical and software engineers. Therefore, this finding emphasizes the contagiousness of TD effects in the life cycle. Disciplines in later phases of the life cycle are more affected by TD. We could identify factors that influence practitioners in mechanical engineering preventing initiating TD incidents (cp. Fig. 4). This information can be elaborated on when designing a mechatronic TD management framework.

### 5.2    Results Compared to Previous Studies

*Agreements with previous work* – In a study of factors that lead development teams to incur TD in Software Projects, Rios et al. [5] concludes that: "[…] Of these, the most often mentioned were deadlines (described in 17 different TD instances) […]". Furthermore, "[…] We can see that the categories Planning and Management and Lack of Knowledge are the most cited […] distributed across different types of debt". In this study, "Lack Of Time/Time To Market/Time To Customer" is considered to be the most frequent TD cause. In addition, "Planning And Management" represent the most frequent TD cause cluster. Besker et al. [10] stated that: "[…] Complex Architectural Design and Requirement TD generate the most negative impact on daily software development work, both when studying the mean and the median values. Regarding the impact of TD types in daily software development work, the significant impact of Architecture TD and Requirements TD can be observed in this study as well. Furthermore, Besker et al. [17] acknowledges that "In both startups and mature organizations, there is often a peak of accumulated TD at the beginning of feature development. "Although in this study, only medium and large-sized companies are represented, Requirements TD and Architecture TD are two of the most frequently mentioned types with the largest impacts. These two TD types are the first two possible TD types in the life cycle of a mechatronic product. Thus, an accumulation of TD at the beginning of the life cycle can be confirmed in this study and in *TDinMechatronics*. Dong & Vogel-Heuser [18] stated that: "Industrial case studies show TD from the mechanical discipline has significant interest for the software discipline of automated production systems. "This study confirmed the postponed accumulation of TD from mechanical to software engineering.

*Controversy to previous work* – Ampatzoglou et al. [7] stated that: "[…] the most frequently identified TD type is test debt. It is followed by code debt and architectural debt. […] Next, documentation, design, and infrastructure debt are found in many cases (55-60%). Requirements, build, and versioning debt seem to be the least commonly identified types of TD. "In this study, the frequency of Test TD is considered to be slightly below average. Architecture TD and Requirements TD are the most frequent TD types, whereas Code TD and Versioning TD seem to occur rarely in mechatronics. These conclusions above apply for all experts, including the group of experts in software and interdisciplinary development. Besker et al. [10] concluded that: "[…] it seems that TD would have a major impact on the extra costs (interest) due to Requirement TD and Testing TD, while Documentation, Architecture (complex and lack of reusability), and Code TD have been reported as quite important as well. "Regarding the impact of TD types in this study, Requirements TD generates the most significant impact among all TD types, followed by Variants TD, Infrastructure TD, and Architecture TD. On the contrary, Versioning TD and Code TD are not considered to be impactful TD types in this study. Codabux & Williams [19] addressed the customer requests as being: "[a] predominant factor in determining if there are available resources to address TD. The product owner prioritizes TD based on customer needs. "In this study, "Customer Requests "are classified as a TD symptom. However, Customer Requests and other TD symptoms related to the customer, such as "Rejected by Costumer ", play a minor role in comparison with other TD symptoms. These customer-orientated TD symptoms may be impactful, but they occur exceptionally rarely according to the experts of this sample. The reason for the differences identified may originate from the different settings of the study, or this sample of experts may have assimilated the subject of TD differently.

### 5.3 Threats to Validity

The four aspects of validity [11, 16] are discussed in detail in the context of the threats to validity identified concerning our study, which may jeopardize the validity of the results. More than one researcher was involved in the design and analysis of the work. The member checking method [11] was applied to increase the validity of the questionnaire. One research colleague confirmed the formulation of the questions. The full online questionnaire is available and may be used to replicate this study. Construct validity refers to the degree to which selected measures represent the research questions to be investigated. In this study, the experts already have a fundamental understanding of TD from their previous interviews for *TDinMechatronics*. The experts received the previous results with the online questionnaire so that they can consolidate and expand their understanding of TD. Additionally, specific definitions related to TD characteristics are clarified in detail in the questionnaire. Internal validity refers to the accuracy of inferences and might be of concern, as the range of selected companies does not show a completely even distribution within their respective industries. Besides, we contacted all 50 experts; nevertheless, only one expert in the electrical and software engineering domain (EE + SE) responded. In order to address this bias, analysis of correlations concerning the disciplines are all marked with the specific discipline description, and multiple disciplines are excluded from the isolated analysis. External validity refers to the capacity to generalize the findings. This study covers a sample of 15 selected questionnaires from industry, which merely gives limited generalizability. Yet, the comparison of the 15 questionnaires to the answers of the same samples in the expert interview offer great insights, at which point the TD characteristics diverge and reveal evidence for future TDM. Reliability is concerned with the extent to which the specific researchers influence the data, the analysis, and the results. Three researchers of the institute were involved in the design and analysis of the work. The peer debriefing and member checking methods [11] were applied to improve the validity of a case study.

## 6 Conclusion and Outlook

After the first case study with 50 expert interviews, this empirical research expands the picture of TD in mechatronic companies substantially. This study gives an insight into how mechatronic companies understand, perceive, evaluate, track, avoid, and manage TD. Based on the knowledge from the previous study of *TDinMechatronics*, 15 experts from the same sample evaluated 14 TD types, 47 TD causes, and 33 TD symptoms regarding their frequency and impact. Although some characteristics overlap with the previous results, the frequency and impact evaluation results show that TD characteristics, which were confined to the background in the first study, became dominant in the general and domain-specific considerations in this study (RQ1). Comparing the TD types, Requirements TD and Infrastructure TD remain the most frequently occurring TD types in both studies. However, the high values of Variants TD in terms of its frequency and impact, and Manufacturing TD in terms of its impact are especially evident outcomes of this study. Especially, we have identified divergent foci regarding software engineers working in mechatronics (e.g., in embedded software). Software engineers, whose previous research focuses mainly on Code TD, Test TD, and Architecture TD, voted in mechatronics for high values in Requirements TD, Variants TD, and Infrastructure TD. All three are of the highest frequency and largest impact. Each discipline focuses on particular TD characteristics. Whether working in cooperation or individually, discipline-specific strategies must be developed. Therefore, prior research results in software engineering should be examined carefully before building on alleged synergies. Also regarding the TD causes, besides the most evident ones, "Save Costs" and "Lack Of Time/Time To Market/Time To Customer", software engineers report "No Holistic View (of the mechatronic product) ", "Great Complexity", "Due to Management Decision", and "One Discipline is More Established in the Industry" as some of the ones with the highest frequency and impact. Concerning the TD indicators, the diverging results of *TDinMechatronics* and this study are especially large. TD indicators, e.g., "One Discipline Decides For Another One", "No Transparent Decision Finding Process", "To-Dos are Unclear" are revealed to be challenging for the company (RQ2). Furthermore, the average frequency and impact of the indicators that are listed increasingly from mechanical (lowest) through electrical to software engineering (highest) show that more attention also needs to be paid to this group of software engineers working in an interdisciplinary environment (RQ3). Of the five elaborated points, "Large Output Quantity "and "Uncertain Market Condition", followed by Mechatronic Development", are the most important factors in preventing TD (RQ4).

The future research of TD in mechatronics, and its management, should focus on the automated identification of TD incidents. The classification concept of active and passive TD types that either directly induce a TD incident exposure or passively exist in the process and induce outbreaks in later stages, or even in later life cycles (e.g., at reuse or when applying the copy-paste-modify strategy), are to be investigated in detail. Besides the results presented, the raw material of the questionnaires allows an in-detail correlative investigation of the TD incidents found by the first study: leading position of the expert, TD type, TD cause, TD symptoms, TD consequence. These are not presented in this paper, as this would exceed the framework of the work. Theoretical research should elaborate on patterns in the cause-symptom-consequence relations of TD incidents in order to limit and prevent TD incidents before they occur and keep them from spreading. TD in mechatronic systems requires more attention from researchers and practitioners if they are to systematically integrate the TDM mechanisms and thinking into organizational systems.

## ACKNOWLEDGMENTS
The authors kindly thank the participating experts and companies for their support with informative industrial insights, as well as for their trust and openness during the study.